\documentclass[11pt,reqno]{amsart}
\usepackage{epsfig}
\usepackage{graphicx}
\usepackage[numbers,sort&compress]{natbib}

\begin{document}
\baselineskip=0.6cm 
\theoremstyle{plain}
\title{Monopole blocking governed by a modified KdV type equation}
\author{X. Y. Tang$^{1}$, J. Zhao $^{2}$, F. Huang$^{2}$, and S. Y. Lou$^{1,3}$}
\dedicatory{
$^{1}$ Department of Physics, Shanghai Jiao Tong University, Shanghai, 200030, China\\
$^{2}$ Meteorology, Ocean University of China, Qingdao, 266003, P. R. China\\
$^{3}$ Department of Physics, Ningbo University, Ningbo 315211, P.
R. China}

\date{29 Nov. 2008}

\begin{abstract}
A type of coupled variable coefficient modified Korteweg-de Vries
system is derived from a two-layered fluid system. It is known that
the formation, maintenance, and collapse of an atmospheric blocking
are always related with large-scale weather or shorts term climate
anomalies. One special analytical solution of the obtained system
successfully features the evolution cycle of an atmospheric monopole
type blocking event. In particular, our theoretical results captures
a real monopole type blocking case happened during 19 Feb 2008 to 26
Feb 2008 can be well described by our analytical solution.
\end{abstract}

\maketitle

\section{Introduction}
Atmospheric blocking is a fundamental large-scale weather phenomena
in mid-high latitudes in the atmosphere that has a profound effect
on local and regional climates. The life cycle of an atmospheric
blocking always brings about large-scale weather or short terms
climate anomalies. Therefore, it is rather important to predict an
atmospheric blocking in regional midterm weather forecast and
short-term climate trend prediction.

There are three types of patterns for an atmospheric blocking
anticyclone, i. e., monopole type blocking (or omega type blocking),
dipole type blocking, and multi-pole type blocking. During the past
many years, the dipole type blocking has been studied a lot since
its first discovery by Rex \cite{rex}. Malguzzi and
Malanotte-Rizzoli first used the Korteweg de-Vries (KdV) Rossby
soliton theory to study dipole type blocking. While unfortunately
their analytical results failed to describe the onset, developing,
and decay of a blocking system. In fact, the important atmospheric
blocking can be explained by many different theories except for KdV
type equations. Recently, Luo et al. proposed the envelope Rossby
soliton theory based on the deduced nonlinear Schr\"odinger (NLS)
type equations and successfully explained the blocking life cycle
numerically. More recently, we have found that variable coefficient
KdV equation can analytically features the life cycle of a dipole
blocking if introducing a time-dependent background field
\cite{htl}. In addition, it has been revealed in Ref. \cite{htl}
that none zero boundary values and time-dependent background
westerly are vital factors to explain a life cycle of a dipole type
blocking.

In this paper, we are motivated to introduce time into the
background flow and boundary conditions to derive new types of
equations from a two-layered fluid to investigate atmospheric
blocking systems. The paper is organized as follows. In Section 2, a
type of coupled variable coefficient modified KdV type system is
derived from a two-layered fluid model. Then we give a special
analytical solution with many arbitrary functions and constants in
Section 3. In Section 4, we assume some special values of the
parameters in the analytical solution to obtain an approximate
analytical expression for the stream functions which can describe a
typical monopole type blocking event. The whole life cycle of the
monopole blocking is graphically displayed, which really captures
the feature of a real observational monopole blocking case happened
in the 2008 snow storm in China. Last section is a short summary and
discussion.

\section{Derivation of the coupled variable coefficient modified KdV type system}
The starting two-layered fluid model is
\begin{eqnarray}
&&q_{1t}+J\{\psi_1,q_1\}+\beta\psi_{1x}=0,\label{eq1}\\
&&q_{2t}+J\{\psi_2,q_2\}+\beta\psi_{2x}=0,\label{eq2}
\end{eqnarray}
where
\begin{eqnarray}
&&q_1=\psi_{1xx}+\psi_{1yy}+F(\psi_2-\psi_1),\label{q1}\\
&&q_2=\psi_{2xx}+\psi_{2yy}+F(\psi_1-\psi_2),\label{q2}
\end{eqnarray}
and $J\{a,b\}\equiv a_xb_y-a_yb_x$. In Eqs. \eqref{eq1}--\eqref{q2},
$F$ is a weak coupling constant between two layers of the fluid and
$\beta=\beta_0(L^2/U)$, $\beta_0=2(\omega_0/a_0)\cos(\phi_0)$, where
$a_0$ is the earth's radius, $\omega_0$ is the angular frequency of
the earth's rotation and $\phi_0$ is the latitude, $U$ is the
characteristic velocity scale. The derivation of the dimensionless
equations \eqref{eq1} and \eqref{eq2} is based on the characteristic
horizontal length scale $L = 10^6$m and the characteristic
horizontal velocity scale $U=10$m/s \cite{ped}. A type of coupled
KdV equations has been derived from the system
\eqref{eq1}-\eqref{q2}, and its Painlev\'e property and soliton
solutions have also been discussed \cite{tong1,tong2}.

It is a common treatment to rewrite the stream function  $\psi$ by
two parts, namely, $\psi=\psi_{0}(y,t)+\psi'$ with $\psi_{0}$ being
the background flow term, introduce the stretched variables
$\xi=\epsilon(x-c_0t),\tau=\epsilon^3t$, ($c_0$ is a constant), and
take the background field $\psi_{0}$ only as a linear function of
$y$, and then expand the stream function as
$\psi'=\sum_{n=1}^{\infty}\epsilon^n\psi'_{n}(\xi,y,\tau)$.
Recently, considering the fact that background flow and the shear of
the flow both could be time dependent, a new treatment was adopted
in Refs. \cite{htl} to view background field $\psi_0$ as a function
arbitrarily depending on $(y,t)$ and then further expand it as
$\psi_0=U_0(y)+\sum_{n=1}^{\infty}\epsilon^nU_n(y,\tau)$.

Below, a new type of coupled variable coefficient modified KdV type
system is derived in a different way from that used in Ref.
\cite{tong1}. First, we introduce a variable transformation
$x'=a_{11}x+a_{12}y,y'=a_{22}y$, and denote
$a_{22}^2=c_1,a_{11}a_{22}=c_2,a_{11}^2+a_{12}^2=c_3$. Then, we
rewrite the stream functions
$\psi_i=\psi_{i0}(y',\tau)+\psi'_i(\xi,y',\tau)(i=1,2)$ with the
stretched variables $\xi=\epsilon(x'-c_0t),\tau=\epsilon^3t$, ($c_0$
is a constant), and finally make the expansions
$\psi_{10}=V_0(y',\tau)+\sum_{n=1}^{\infty}\epsilon^nV_n(y',\tau),\psi_{20}=U_0(y',\tau)+\sum_{n=1}^{\infty}\epsilon^nU_n(y',\tau)$,
and $\psi_i'=\sum_{n=1}^{\infty}\epsilon^n\psi'_{in}(\xi,y',\tau)$.
In addition, we introduce $F = F_0\epsilon, \beta =
\beta_0\epsilon^2$. Then we substitute all the expansions into Eqs.
\eqref{eq1} and \eqref{eq2} with Eqs. \eqref{q1} and \eqref{q2}, and
then vanish all the coefficients of each order of $\epsilon$. For
notation simplicity, the primes are dropped out in the following.

In the first order of $\epsilon$, we obtain
\begin{eqnarray}
\psi_{11}&=&(C_0(\tau)\xi+A_1(\xi,\tau))B_1(y,\tau)\equiv
(C_0\xi+A_1)B_1,\label{psi11}\\
\psi_{21}&=&(C_1(\tau)\xi+A_2(\xi,\tau))B_2(y,\tau)\equiv
(C_1\xi+A_2)B_2,\label{psi21}
\end{eqnarray}
where $B_1$ and $B_2$ satisfy
\begin{eqnarray}
(c_0+c_2V_{0y})B_{1yy}-c_2V_{0yyy}B_1=0,\label{B1}
\end{eqnarray}
and
\begin{eqnarray}
(c_0+c_2U_{0y})B_{2yy}-c_2U_{0yyy}B_2=0,\label{B2}
\end{eqnarray}
respectively. The general solutions of Eqs. \eqref{B1} and
\eqref{B2} read
\begin{eqnarray}
B_1=\left(F_1\int (c_0+c_1V_{0y})^{-2}{\rm
d}y+F_2\right)(c_0+c_1V_{0y}),\label{reB1}
\end{eqnarray}
and
\begin{eqnarray}
B_2=\left(F_3\int (c_0+c_1U_{0y})^{-2}{\rm
d}y+F_4\right)(c_0+c_1U_{0y}),\label{reB2}
\end{eqnarray}
where $F_i,(i=1,2,3,4)$ are arbitrary integration functions of
$\tau$.

In the second order of $\epsilon$, we have
\begin{eqnarray}
\psi_{12}&=& B_3A_{1\xi}+B_4A_2+
B_5A_1^2+B_6\xi A_1+B_7A_1+B_8\xi^2+B_9\xi,\label{psi12}\\
\psi_{22}&=&B_{11}A_{2\xi}+B_{12}A_1+B_{13} A_2^2+B_{14}\xi
A_2+B_{15}A_2+B_{16}\xi^2+B_{10}\xi,\label{psi22}
\end{eqnarray}
where $B_i\equiv B_{i}(y,\tau),(i=3,4,...,16)$ are determined by a
system of equations presented in Appendix A.

In the third order of $\epsilon$, if we assume
\begin{eqnarray}
&&\psi_{13}=B_{19}A_{2\xi}+B_{20}A_{1\xi\xi}+(B_{21}A_1+B_{22}\xi+B_{23})A_{1\xi}+B_{24}A_1^3+B_{25}A_1^2+B_{30}A_2^2\nonumber\\
&&\quad+(B_{26}A_2+B_{27}\xi^2+B_{28}+B_{29}\xi)A_1+(B_{31}\xi+B_{32})A_2+B_{33}\xi^3+B_{34}\xi+B_{35}\xi^2\nonumber\\
&&\quad +\int
B_{36}A_{1\xi}^2+(B_{37}A_2+B_{38}\xi A_1)A_{1\xi}+B_{39}A_1^2+(B_{40}\xi+B_{41})A_1+B_{42}A_2 {\rm d}\xi,\label{psi13}\\
&&\psi_{23}=B_{43}A_{1\xi}+B_{44}A_{2\xi\xi}+(B_{45}A_2+B_{46}\xi+B_{47})A_{2\xi}+B_{48}A_2^3+B_{49}A_2^2+B_{54}A_1^2\nonumber\\
&&\quad +(B_{50}A_1+B_{51}\xi^2+B_{52}+B_{53}\xi)A_2+(B_{55}\xi+B_{56})A_1+B_{57}\xi^3+B_{58}\xi+B_{59}\xi^2\nonumber\\
&&\quad +\int B_{60}A_{2\xi}^2+(B_{61}A_1+B_{62}\xi
A_2)A_{2\xi}+B_{63}A_2^2+(B_{64}\xi+B_{65})A_2+B_{66}A_1{\rm
d}\xi,\label{psi23}
\end{eqnarray}
and requiring $B_i\equiv B_{i}(y,\tau),(i=19,20,...,66)$ satisfy a
system of equations (we do not write them down here for they are too
long while can be easily retrieved following the above procedures),
then we arrive at a coupled variable coefficient modified KdV type
system
\begin{eqnarray}
&&A_{1\tau}+e_{16}A_{1\xi\xi\xi}+(e_{45}A_1^2+(e_{42}\xi+e_{24})
A_1+e_{32}\xi^2+e_{30}+e_{31}\xi+e_{12}A_2)A_{1\xi}+e_{44}A_{1\xi}^2\nonumber\\
&&+e_{36}A_1^2+(e_{14}+e_{15}\xi)A_1+e_{21}A_{2\xi\xi}+(e_{18}A_1+e_{20}A_2+e_{22}\xi+e_{23})A_{2\xi}\nonumber\\
&&\quad +(e_{29}+e_{17}A_1+e_{28}\xi)
A_{1\xi\xi}+e_{38}\xi+e_{39}\xi^2+e_{37}+e_{13}A_2=0,\label{red1}\\
&&A_{2\tau}+e_{25}A_{2\xi\xi\xi}+(e_{35}A_1+e_{34}A_2^2+(e_{6}+e_7\xi)A_2+e_1+e_2\xi
+e_3\xi^2)A_{2\xi}+e_{33}A_{2\xi}^2\nonumber\\
&&\quad+e_8A_2^2+(e_5+e_4\xi)A_2+e_{40}A_{1\xi\xi}+(e_{11}\xi+e_{26}+e_{19}A_2+e_{41}A_1)A_{1\xi}\nonumber\\
&&\quad+(e_9\xi+e_{10}+e_{27}A_2)A_{2\xi\xi}+e_{43}A_1+e_{47}\xi+e_{48}\xi^2+e_{46}=0,\label{red2}
\end{eqnarray}
with $e_i\equiv e_i(\tau),(i=1,2,...,48)$ being arbitrary functions
of the indicated variable.

\section{Special exact solutions}
Since equations \eqref{red1}-\eqref{red2} constitute a coupled
variable coefficient nonlinear system, it is not easy to obtain its
general solution. Here we present a quite special solution of the
system. It is easy to see that if we suppose $A_1=aA_2$ with
constant $a$, then Eqs. \eqref{red1} and \eqref{red2} degenerate to
one modified KdV type equation
\begin{eqnarray}
&&A_{2\tau}+e_{16}A_{2\xi\xi\xi}+(m_{10}+a e_{17}A_2+e_{28}\xi)A_{2\xi\xi}+ae_{44}A_{2\xi}^2+ae_{36}A_2^2+(m_5+e_4\xi) A_2\nonumber\\
&&~~+(a^2e_{45}A_2^2+(ae_{42}\xi+m_6)A_2+e_{32}\xi^2+m_2\xi+m_1)A_{2\xi}+e_{48}\xi^2+e_{47}\xi+e_{46}=0,\label{mkdv}
\end{eqnarray}
where $m_i\equiv m_i(\tau),(i=1,2,5,6,10)$ are given by $
m_1=e_1+ae_{26},m_2=e_2+ae_{11},m_5=e_5+ae_{43},
m_6=e_6+ae_{35}+a^2e_{41}+ae_{19},m_{10}=e_{10}+ae_{40},$ when $e_3
= e_{32} ,e_9 = e_{28}  ,e_{15} = e_4,e_{34}= a^2 e_{45} , e_{33}  =
ae_{44}, e_7  = a e_{42} , e_{37} =ae_{46}, e_{38} =a e_{47}, e_{27}
= a e_{17} , e_{39}  =a e_{48}, e_8 = ae_{36}, e_{25}  = e_{16} ,
e_{21} = a e_{10}+a^2 e_{40} -a e_{29}, e_{13}  = -a e_{14}
+ae_5+a^2 e_{43},e_{23} = -a e_{30} +a e_1+a^2 e_{26} , e_{22} = a^2
e_{11} -a e_{31} +a e_2,e_{20} = -a e_{18}-a e_{12} -a^2 e_{24} +a
e_6+a^2 e_{19}+ a^2e_{35}+a^3 e_{41}. $

Further, it can be verified that the modified KdV equation
\eqref{mkdv} can be transformed to the standard one
\begin{eqnarray}
P_T+6P^2P_X+P_{XXX}=0.\label{smkdv}
\end{eqnarray}
if
\begin{eqnarray}
A_2=-\frac{m_6f_1f_2^2}{2a^2f_{4\tau}}+f_2P(X,T),\label{A21}
\end{eqnarray}
where $T\equiv f_4(\tau)$, $X$ is given by
\begin{eqnarray}
X=\frac{6f_1}{a^2}\xi-\frac{3f_3}{2a^4},\label{XT}
\end{eqnarray}
and $f_i\equiv f_i(\tau),(i=1,2,3,4)$ are arbitrary functions, with
conditions
\begin{eqnarray}
&&m_1=\frac{m_6^2f_1+e_{45}f_{3\tau}}{4a^2f_1e_{45}},
m_2=-\frac{f_{1\tau}}{f_1}, m_5= -\frac{f_{2\tau}}{f_2},e_{16}=\frac{a^6f_{4\tau}}{216f_1^3},\\
&&e_{45}=\frac{f_{4\tau}}{f_1f_2^2}, \quad 2a^2e_{45}^2e_{46}-m_5m_6e_{45}-m_{6\tau}e_{45}+m_6e_{45\tau}=0,\\
&&e_{17}=e_{28}=e_{32}=e_{36}=e_4= e_{42}=e_{44}=e_{47}=
e_{48}=m_{10}= 0.
\end{eqnarray}
Hence, it is easy to obtain exact solutions of Eq. \eqref{mkdv}
based on the solutions of Eq. \eqref{smkdv} through Eq. \eqref{A21}
with Eq. \eqref{XT}. One classical soliton solution of the mKdV
equation \eqref{smkdv} is $P=K{\rm sech}(KX-K^3T)$ with a constant
$K$. Now just using this typical solution, we can easily write down
a special solution of the original system \eqref{eq1}-\eqref{eq2} as
\begin{eqnarray}
&&\psi_1\approx V_0(a_{22}y,\tau)+\epsilon
V_1(a_{22}y,\tau)+\epsilon
B_1(a_{22}y,\tau)\left\{C_0\epsilon(a_{11}x+a_{12}y-c_0t)
-\frac{m_6f_1f_2^2}{2af_{4\tau}}\right.\nonumber\\
&&\qquad \left.+af_2K{\rm
sech}\left[\frac{6Kf_1\epsilon}{a^2}(a_{11}x+a_{12}y-c_0t)-\frac{3Kf_3}{2a^4}
-K^3f_4\right]\right\},\label{repsi1}\\
&&\psi_2\approx U_0(a_{22}y,\tau)+\epsilon
U_1(a_{22}y,\tau)+\epsilon
B_2(a_{22}y,\tau)\left\{C_1\epsilon(a_{11}x+a_{12}y-c_0t)
-\frac{m_6f_1f_2^2}{2a^2f_{4\tau}}\right.\nonumber\\
&&\qquad \left.+f_2K{\rm
sech}\left[\frac{6Kf_1\epsilon}{\beta_0^2a^2}(a_{11}x+a_{12}y-c_0t)-\frac{3Kf_3}{2a^4}
-K^3f_4\right]\right\},\label{repsi2}
\end{eqnarray}
with $\tau=\epsilon^3 t$, $B_1$ and $B_2$ determined by Eqs.
\eqref{reB1} and \eqref{reB2}, respectively. From the analytical
solution of stream functions $\psi_1$ and $\psi_2$, it is easy to
derive the corresponding background westerly flows
$\bar{u}_1=-\partial \psi_{10}/\partial y=-\partial
V_0(y,\tau)/\partial y-\epsilon\partial V_1(y,\tau)/\partial y$, and
$\bar{u}_2=-\partial \psi_{20}/\partial y=-\partial
U_0(y,\tau)/\partial y-\epsilon\partial U_1(y,\tau)/\partial y$,
which are all left as arbitrary functions of the indicated
variables.

\section{Analytical diagnosis}
\begin{figure}[htbp]
\includegraphics[width=4.5cm]{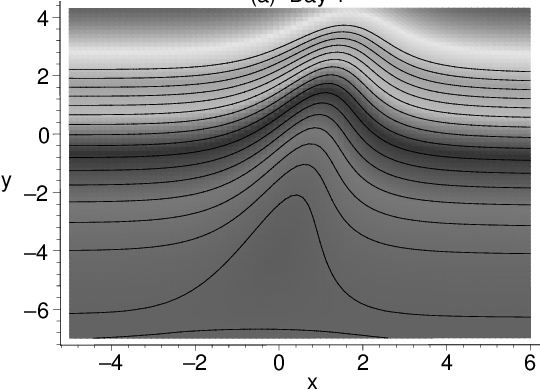}
\includegraphics[width=4.5cm]{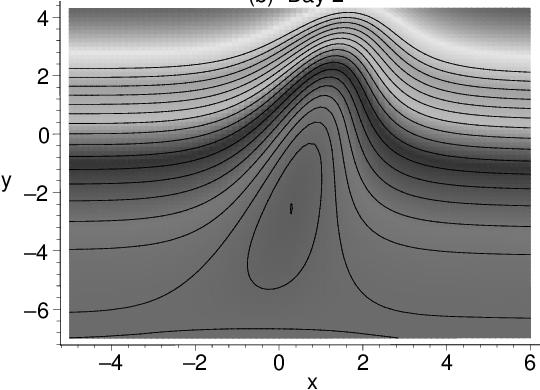}
\includegraphics[width=4.5cm]{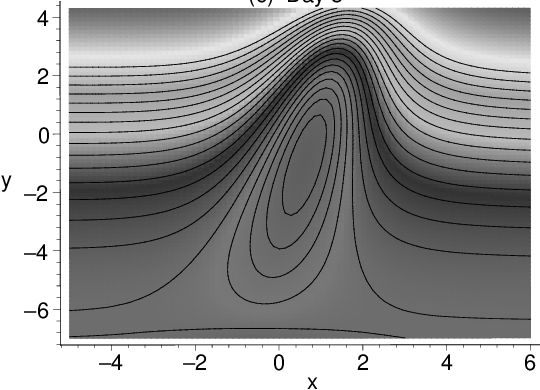}\\
\includegraphics[width=4.5cm]{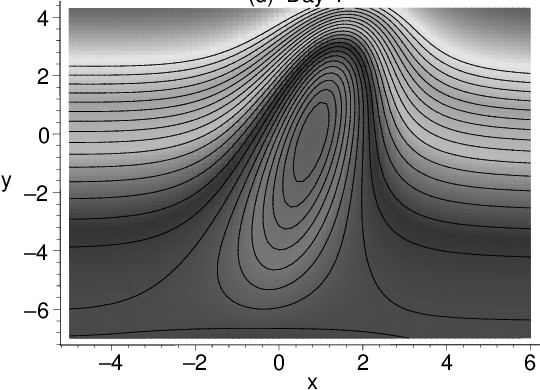}
\includegraphics[width=4.5cm]{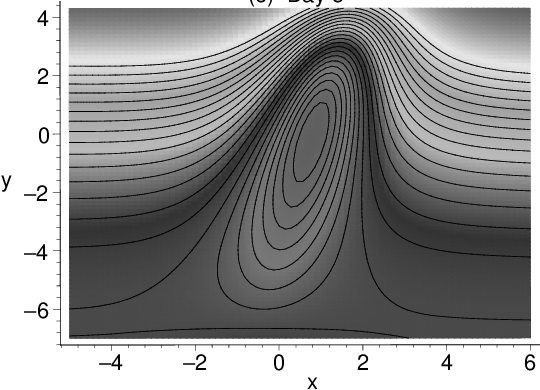}
\includegraphics[width=4.5cm]{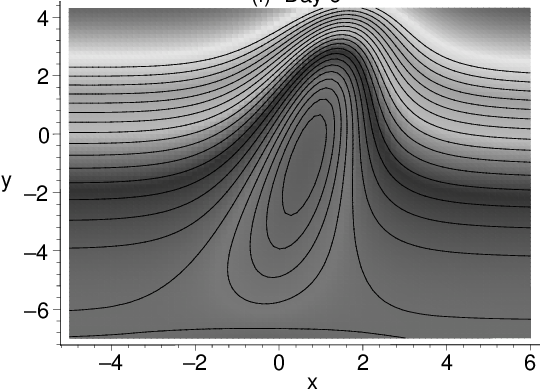}\\
\includegraphics[width=4.5cm]{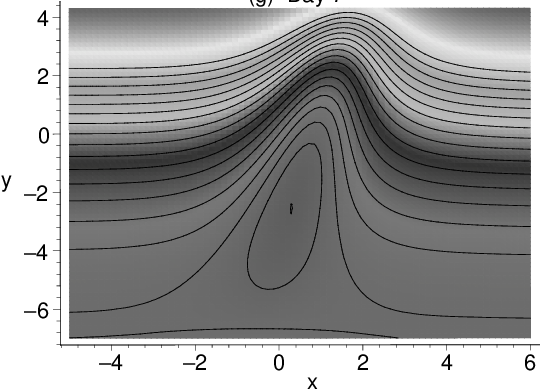}
\includegraphics[width=4.5cm]{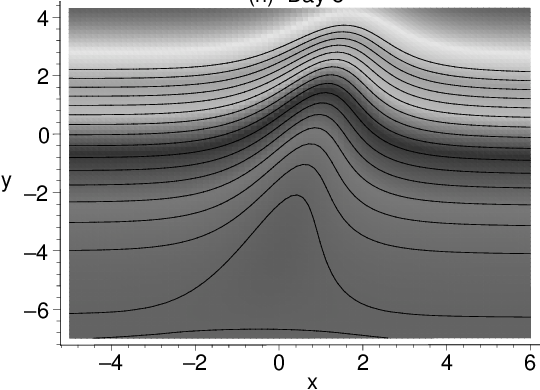}
\caption{A monopole type blocking life cycle from the theoretical
solution \eqref{repsi1} with the parameters given by \eqref{cons}.
The contour interval (CI)=0.12.\label{fig}}
\end{figure}

By selecting the arbitrary functions and constants appropriately,
the approximate analytical solution \eqref{repsi1}-\eqref{repsi2}
can be responsible for different kinds of atmospheric blocking
phenomenon. Here we give a typical example when the functions and
constants are taken as
\begin{eqnarray}
&& V_0=-0.006y^2+0.001y-3{\rm
sech}(0.25y-1.25),\quad a=C_0=f_1=f_2=a_{22}=1,\nonumber\\
&&c_0 = m_6=F_1=V_1=0,\quad f_3=-24f_4(\tau)+0.1,\quad F_2={\rm
sech}(0.5(t-4.6)),\label{cons}
\\ &&a_{11} = 0.3,\quad a_{12}=-0.06,\quad c_1=10,\quad \epsilon=0.1,\quad K=-6.\nonumber
\end{eqnarray}
In this case, the solution \eqref{repsi1} can describe a monopole
type blocking event as depicted in Figure \ref{fig}. Considering the
fact that the basic flow, mainly the basic westerly and the shear of
the basic westerly, plays a significant role on the blocking
developing process \cite{htl,rex,Mal,luobook}, it is hence
reasonable to introduce the $y^2$ related term.

Evidently, a whole life cycle of a monopole type blocking event,
namely, the onset, development, maintenance, and decay processes,
are clearly presented in Figure \ref{fig}. The streamlines are
gradually deformed, and the anticyclonic high in the north develops
at the second day (Fig. 1a). It is strengthened daily. At around the
fourth day (Fig. 1e), it is at its strongest stage and then become
weaker and eventually disappear after the seventh day (Figs. 1e-h).
Obviously, Fig. 1 possesses the phenomenon's salient features
including their spatial-scale and structure, amplitude, life cycle,
and duration. Therefore, Fig. 1 is a very typical monopole type
blocking episode in one layer of the fluid.

\begin{figure}[htbp]
\includegraphics[width=12cm]{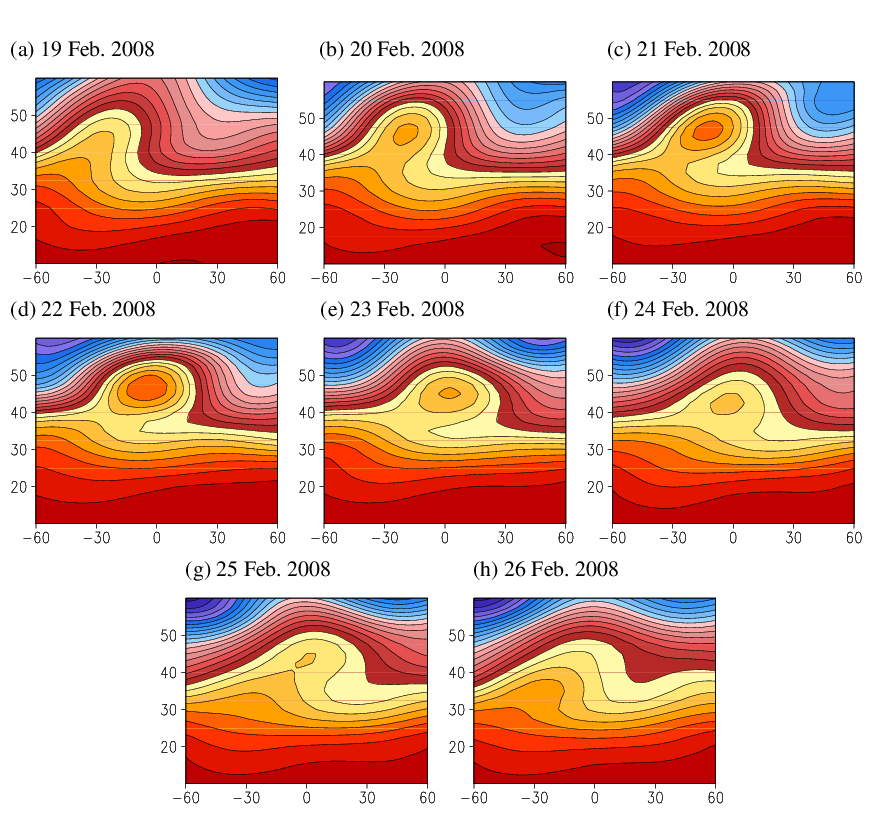}
\caption{Filtered geopotential height at 300-hPa pressure level of a
monopole blocking case during 19-26 Feb 2008. The $x$-axis is
longitude, and the $y$-axis is latitude. Contour interval is 6
gpdm.\label{fig2}}
\end{figure}

A real observational blocking case happened during 19 Feb 2008 to 26
Feb 2008 is shown in Fig. \ref{fig2} appearing a monopole pattern.
It is easily found that the life cycle of this blocking lasts about
eight days experiencing three stages: onset (19-20 Feb, 2008),
mature (21-22 Feb, 2008) and decay (23-26 Feb, 2008) periods, which
resembles those displayed in Figure 1.

\section{Summary and discussion}
Considering a time-dependent basic westerly, we have derived a type
of coupled variable coefficient modified KdV type system from a
two-layered fluid model, with the time dependent coefficients
resulted from the time dependent basic flows and time dependent
boundary conditions. Instead of possessing a linear meridional
shear, the mean flow is assumed to be a combination of a quadratic
function of $y$ and ${\rm sech}(y)$. The boundary conditions remain
as unknown functions with some complicated relations. Under a set of
parameters, our analytical solution nicely described a real
observational blocking life cycle from 19 Feb to 26 Feb 2008,
indicating the onset, mature and decay phases in its developing
process in one of the layers of the fluid. In the other layer,
similar or different types of blocking can be found when setting the
unknown parameters.

Therefore, it is revealed that blocking can also be governed by
modified KdV type system and its analytical solutions can also
features the life cycle of a blocking. It is worth further
investigations on how variations of the time-dependent background
westerlies influence the type of a blocking and the evolution of a
blocking during its life period since the role of the weak
westerlies on a blocking is still unclear.

\section {Acknowledgement}
The work was supported by the National Natural Science Foundation of
China (No. 10735030, No. 10547124, No. 90503006, and No. 40305009),
National Basic Research Program of China (973 program) (No.
2007CB814800 and No. 2005CB422301),Specialized Research Fund for the
Doctoral Program of Higher Education (20070248120), Program for
Changjiang Scholars and Innovative Research Team in University
(PCSIRT No. IRT0734),  and Program for New Century Excellent Talents
in University (NCET).

\section*{Appendix A}
$B_i\equiv B_{i}(y,\tau),(i=3,4,...,18)$ are determined by the
following equations
\begin{eqnarray}
&&c_1(c_0+c_2V_{0y})B_{3yy}-c_2a_{22}^2V_{0yyy}B_{3}+2a_{12}( c_1
a_{11} V_{0y}+c_0a_{22})B_{1y}=0,
\end{eqnarray}
\begin{eqnarray}
&&c_1(c_0+c_2V_{0y})B_{4yy}-c_1c_2V_{0yyy}B_4+(c_0+c_2V_{0y})F_0
B_2=0,
\end{eqnarray}
\begin{eqnarray}
&&2(c_1c_2^3 V_{0y}^3+3  c_0c_1c_2^2 V_{0y}^2 +
c_0^3c_1+3c_0^2c_1c_2 V_{0y})B_{5yy}+c_1c_2^3B_{1}^2 V_{0yyy} V_{0yy}\nonumber\\
&&\quad -2c_1c_2(2c_0c_2V_{0y}+c_0^2+c_2^2V_{0y}^2)
V_{0yyy}B_5-c_2^2(c_0+c_1c_2V_{0y}) V_{0yyyy}B_{1}^2=0,
\end{eqnarray}
\begin{eqnarray}
&&(c_{1}c_{2}^3 V_{0y}^3+3  c_0c_{1}c_{2}^2 V_{0y}^2 +
c_0^3c_1+3c_0^2 c_{1}^2c_{2}^2 V_{0y} )B_{6yy}-c_1c_{2}^2(c_0
C_0+c_{2}C_0 V_{0y}) V_{0yyyy}B_{1}^2
\nonumber\\
&&\quad +c_1c_{2}c_{2}^2 B_{1}^2 C_0 V_{0yyy}
V_{0yy}-c_1c_{2}(c_{2}^2 V_{0y}^2 +2c_0 c_{2} V_{0y}+c_0^2) V_{0yyy}
B_6=0,
\end{eqnarray}
\begin{eqnarray}
&&(V_{0y}^2 a_{11}^2 a_{22}^6+2 c_1 V_{0y} a_{22}^3 a_{11} c_0+c_1^2
c_0^2)B_{7yy}-c_1^2c_{2}(c_0+c_{2}V_{0y})V_{0yyy}
B_7+a_{11}^2 a_{22}^6 B_{1} V_{0yyy} V_{1y}\nonumber\\
&&\quad -c_1 (c_0^2+c_0c_2U_{0y}+c_0c_2 V_{0y}+c_{2}^2U_{0y}
V_{0y})F_0 B_{1}-c_1^2 c_2(c_0 +c_2V_{0y})V_{1yyy}B_{1}=0,
\end{eqnarray}
\begin{eqnarray}
&&2c_{1}^2( c_{2}^3 V_{0y}^3+3c_0 c_{2}^2V_{0y}^2
+c_0^3+3c_0^2c_{2}V_{0y})B_{9yy}-2c_1^2c_2(c_0^2+2c_0 c_2V_{0y}+c_2^2V_{0y}^2) V_{0yyy}B_8\nonumber\\
&&\quad -c_1^2c_2^2C_0^2(c_0+c_2V_{0y}) V_{0yyyy} B_{1}^2
+c_1^2c_2^3 B_{1}^2 C_0^2 V_{0yyy} V_{0yy}=0,
\end{eqnarray}
\begin{eqnarray}
&&(V_{0y}^2c_1^2c_2^2+2 c_1 V_{0y} c_0c_1c_2+c_1^2
c_0^2)B_{10yy}-c_1^2c_2C_0(c_0+c_2V_{0y}) V_{1yyy}B_{1}-2c_0 F_0 B_2 C_1 c_1 c_2V_{0y}\nonumber\\
&&\quad -c_1^2c_2(c_0B_9+c_2B_9 V_{0y}-c_2C_0B_{1}V_{1y}) V_{0yyy}
-C_0c_1c_2(c_0 + c_2 V_{0y})F_0 U_{0y}B_{1} \nonumber\\
&&\quad+c_1C_1(c_0^2+c_2^2V_{0y}^2)F_0
B_2-c_0c_1C_0(c_0+c_2V_{0y})F_0B_{1}-c_1^2(c_0+c_2V_{0y})V_{0yy\tau}=0,
\end{eqnarray}
\begin{eqnarray}
c_1( c_0+c_2 U_{0y})B_{11yy}+ 2a_{12}(c_1 a_{11}
U_{0y}+a_{22}c_0)B_{2y}-c_1c_2U_{0yyy} B_{11}=0,
\end{eqnarray}
\begin{eqnarray}
c_1( c_0+c_2 U_{0y})B_{12}-a_{11} a_{22}^3 U_{0yyy} B_{12}+( c_0+
c_2U_{0y})F_0B_{1}=0,
\end{eqnarray}
\begin{eqnarray}
&&2c_1^2(c_2^3U_{0y}^3 +3c_0 c_2^2U_{0y}^2+3c_0^2c_2 U_{0y}
+c_0^3)B_{13yy}++c_1^2c_2^3B_2^2 U_{0yyy} U_{0yy} \nonumber\\
&&\quad -2c_1^2c_2(c_2^2 U_{0y}^2+c_0^2+2c_0c_2
U_{0y})B_{13}U_{0yyy} +c_1^2c_2^2(c_0-c_2U_{0y}) B_2^2 U_{0yyyy}=0,
\end{eqnarray}
\begin{eqnarray}
&&c_1^2(c_2^3U_{0y}^3+3c_0 c_2^2U_{0y}^2+3c_0^2 c_2
U_{0y}+c_0^3)B_{14yy}-c_1^2c_2^2C_1(c_0+c_2U_{0y})B_2^2 U_{0yyyy}\nonumber\\
&&\quad -c_1^2c_2 (c_2^2U_{0y}^2+2 c_0c_2
U_{0y}+c_0^2)B_{14}U_{0yyy}+c_1^2c_2^3 B_2^2 C_1 U_{0yyy} U_{0yy}=0,
\end{eqnarray}
\begin{eqnarray}
&&c_1(c_2^2U_{0y}^2+2c_2 U_{0y} c_0+c_0^2)B_{15yy}-c_1c_2
(c_0+c_2U_{0y})B_2U_{1yyy}-c_2(c_0+c_2U_{0y})F_0V_{0y}B_2
\nonumber\\
&&\quad -c_1c_2(c_0+c_2U_{0y})U_{0yyy} B_{15}
 -c_0( c_0+c_2U_{0y})F_0B_2 -c_1c_2^2
B_2 U_{0yyy} U_{1y}=0,
\end{eqnarray}
\begin{eqnarray}
&&2c_1^2(U_{0y}^3 c_2^3+3 U_{0y}^2c_2^2 c_0+3 U_{0y}
c_2c_0^2+c_0^3)B_{17yy}-c_1^3c_2^2C_1^2(c_0c_2U_{0y})B_2^2U_{0yyyy} \nonumber\\
&&\quad+c_1^4c_2^4 B_2^2 C_1^2 U_{0yyy} U_{0yy}-2c_1^3c_2(2c_2
U_{0y}c_0+c_0^2+c_2^2 U_{0y}^2)U_{0yyy} B_{16} =0,
\end{eqnarray}
\begin{eqnarray}
&&c_1(c_2^2U_{0y}^2 +2c_0c_2 U_{0y}
+c_0^2)B_{18yy}-c_1(c_0+c_2U_{0y})U_{0yy\tau}-c_1c_2C_1
(c_0+c_2U_{0y})B_2 U_{1yyy}\nonumber\\
&&\quad +c_1c_2 (c_0+c_2U_{0y})U_{0yyy} B_{10}+ c_0^2F_0(C_0 B_{1}- C_1B_2)+c_2C_0(c_2U_{0y}+2c_0)F_0B_{1} U_{0y} \nonumber\\
&&\quad -c_2 C_1 (c_0+c_2V_{0y}) F_0U_{0y} B_2 -c_0c_2  F_0 B_2
V_{0y} C_1+c_1c_2^2C_1 B_2 U_{0yyy} U_{1y}=0.
\end{eqnarray}

\end{document}